**A New Method for Calculating the Energies Associated with Particle Reactions**

Borros Arneth, Philipps University Marburg and Justus Liebig University Giessen

Dr. Borros Arneth

Institute for Physics

Philipps University Marburg

Renthof 5

35037 Marburg, Germany

Justus Liebig University Giessen

Feulgenstr. 10-12

35392 Giessen

Germany

e-mail: borros.arneth@staff.uni-marburg.de

**Abstract**

A new method in which the energy and mass of elementary particles can be calculated is presented. Gluon–gluon interactions within a single elementary particle are considered, and the number of possible interactions per particle is determined. This procedure can be formalized and standardized via a newly introduced microcanonical partition function. The possibility of calculating the energetic relationships provides new and more in-depth insights into the reaction possibilities of the particles. In addition, the application of this method of calculating the partition function to the known quarks themselves suggests that they are composed of even more elementary particles. The properties of these particles match the Rishons of the Harari–Shupe preon model. In combination with the Rishon model, the method presented here for calculating the energetic situation of particle reactions provides a profound and new understanding of the processes at an absolute elementary level; this opens new possibilities for the calculation and understanding of particle reactions and might change our understanding of particle physics in fundamental ways.

## 1.0 Introduction

### 1.1 Color Force and Strong Interaction

The color force is a fundamental property of quarks and is characterized by associations between baryons [1]. The strong interaction extends beyond neutrons and protons, resulting in extraordinary properties between nucleons that have yet to be reported. The baryons facilitate strong interactions and residual forces that do not taper off with distance. This process effectively results in the confinement of quarks. Conventionally, naturally occurring energy exists and expedites quark–antiquark pair production [2,3,4]. Owing to the strong nature of the color force carried by the exchange of gluons, the quarks do not undergo separation.

Previous studies have reported that within the confines of a given space, quarks behave as free particles, as the color force exerts minimum pressure at short distances [5]. Stronger forces begin to be realized as further separation occurs. This concept is analogous to the widely held scientific understanding that ordinary mesons composed of conventional quarks are bound closely by color force. This construct has immense scientific importance, provides an in-depth understanding of atomic nuclei, and defines the associations and interactions between neutrons, quarks, and protons [5,6,7]. The mediation of these relationships by gluons forms the foundation of quantum chromodynamics and is commonly used to discuss complex physical phenomena, including chiral symmetry breaking, asymptomatic freedom, and head collar confinement [8, 9].

Gluons act as gauge bosons or exchange particles that facilitate interactions between quarks via strong forces. Essentially, gluons enable the exchange of photons between charged particles via the electromagnetic force in addition to binding them to form hadrons. Bosons are carrier particles with full-integer spin numbers, whereas fermions have odd half-integer spin numbers. Both types of particles act as force carriers, with the latter obeying the Pauli exclusion principle [8,9,10]. Conventionally, the color charge depends on the configuration of the particles. For example, mixing red, blue, and green results in white, with a net charge of zero. Additionally, the force may be negative, causing the emergence of cyan, magenta, and yellow. Fundamentally, the color charges exhibit a force that tends to remain constant despite separation but results in a corresponding increase in energy and subsequent spontaneous production of a quark–antiquark pair [11].

The concept of color force is believed to have been introduced by Oscar Greenberg, who first theorized about the existence of quarks in 1964 [12]. Greenberg documented the phenomenon after an extensive analysis of the Pauli exclusion principle and its relation to the construct. He believed that this theory explained the coexistence of quarks within select hadrons although they exhibit identical quantum states [13], which would otherwise have constituted a theoretical impossibility owing to its violation of the Pauli exclusion principle [14,15,16].

In 1964, Greenberg proposed that quarks were parafermions of order 3, and the corresponding implicit degree of freedom was soon perceived by Han and Nambu and referred to as color [13].

Since 1964, there have been extensive developments in quantum chromodynamics toward an improved understanding of quarks and the theory of color force. Murray Gell-Mann, Heinrich Leutwyler, and Harald Fritsch have made immense contributions. The concept has been revolutionary in enhancing the understanding of quantum electrodynamics owing to the many similarities between the two fields. The strong force has been determined to have minimal behavioral autonomy over the basic subatomic particles constituting atomic nuclei.

The theories of the strong force and other associated aspects of quantum chromodynamics (QCD) were developed to provide an improved understanding of the roles of each aspect of the three-value color charge. Fundamental QCD concepts, including gluons, antiquarks, and quarks, remained relatively obscure prior to the discovery of the color force. Related analyses have been fundamental to the furthering of scholarly discourse as well as inspiring a plethora of other industrial applications, including specialist lighting, such as in theatres and outdoors.

Starting from the well-known connection between the Rydberg constant and the electron mass, we start with the following:

$$E_{Rydberg} = \frac{1}{2} m_e c^2 \alpha^2 \qquad (1)$$

where alpha is the Sommerfeld fine structure constant.

There are several methods and formulas for calculating the Sommerfeld fine structure constant, which is equal to approximately 1/137, by using the mathematical constant pi.

We use the following formula for the Sommerfeld fine structure constant alpha:

$$\alpha = \frac{2 \cdot \sqrt{\sqrt{\left(1 - \frac{1}{3\pi}\right)}}}{27\pi^2} \qquad (2)$$

Next, we calculate the smallest quantum of mass and/or energy according to the Rydberg energy:

$$E_{Rydberg} = \frac{1}{2} m_e c^2 \alpha^2 \qquad (1) \qquad \text{in [eV], this yields}$$

$$E_{Rydberg} = 13.60569 \, eV \qquad (3)$$

$$m_{Rydberg} = 13.60569 \, eV/c^2 \qquad (4)$$

This number is postulated here to be the smallest possible amount of mass and/or energy involved in the crystallization of energy toward mass and in mass formation. It has the smallest amount of binding energy.

From both equations above, we obtain a formula for the electron mass as follows:

$$m_e = \frac{1}{2} \, 3^2 \, (3\pi)^4 \left( \frac{1}{\sqrt{\sqrt{\left(1-\frac{1}{3\pi}\right)}}} \right)^4 m_{Ryd} \tag{5}$$

We consider this equation to be a special case of the following generalization of this formula to obtain the following generalized equation.

$$m_{particle} = 2^a \, 3^b \, (3\pi)^c \left( \frac{1}{\sqrt[8]{\left(1-\frac{1}{3\pi}\right)}} \right)^d m_{Ryd}, \tag{6}$$

with a=-1, b=2, c=4, and d=4 for electron e.

We postulate that we can obtain all other particle masses through different combinations of (a, b, c, and d)-tuples. However, how to interpret this last equation remains unclear.

## 2.0 Results

### 2.1. Summarizing the Feynman Diagrams by Using the Tools of Combinatorics:

**The Feynman-like Ensemble of Binding States within a Single Particle is related to the Partition Function of a Particle.**

A single particle can be formulated as a series of different binding states. This series of different formulations of binding states forms an ensemble comparable to the statistical microcanonical ensembles used in statistical thermodynamics and/or statistical quantum mechanics.

In statistical quantum mechanics, if a system can be subdivided into N subsystems, then the total partition function is given by the product of the partition functions for the individual subsystems. Z and/or Omega is the **microcanonical partition function** (which depends on the multiplicity or number of possible configurations of the system/particle) [17]:

$$Z = \prod_{j=1}^{N} \zeta_j \qquad \text{or} \qquad \Omega = \prod_{j=1}^{N} \omega_j \tag{7a+7b}$$

Accordingly, the probability P is as follows [17]:

$$P = \frac{1}{\Omega} = \frac{1}{\prod_{j=1}^{N} \omega_j} \tag{8}$$

Starting from the Feynman-like ensemble for a complex particle, a single binding form is regarded as a microcanonical subsystem here. There are four possible binding forms: two-quark binding (color and anticolor), three-quark binding (three colors), a 3-vertex gluon interaction, and a 4-vertex gluon interaction.

Within each of these binding-type subsystems, all the remaining subsubseries elements represent different subforms of the corresponding binding type. The elements of these subforms have the same physical properties. Thus, their partition function has a power-function form:

$$\varsigma_j = \varsigma_{jj}{}^n \quad \text{or} \quad \omega_j = \omega_{jj}{}^n \tag{9a+9b}$$

Here, $n$ is the number of subsubseries present in the particle, meaning the number of two-quark interactions, three-quark interactions, 3-vertex gluon interactions, or 4-vertex gluon interactions. where $\varsigma_{jj}$ is the energy of each of the basal binding states, i.e., the basal energy of the two-quark binding state, three-quark binding state, 3-vertex state, or 4-vertex state.

The basal energy $\varsigma_{jj}$ of each individual binding state can be visualized by an arrow in a Feynman-like diagram and corresponds to one single gluon. The basal energy of two-quark binding is represented by two arrows and two gluons, one of which is colorless and one of which is colored. The basal energy of three-quark binding is represented by three arrows and three gluons (r, g, and b). The basal energy of a 3-vertex gluon interaction can be visualized by three half-circles (the Dyson–Schwinger equations and the swordfish diagram) and is given by an arrow of length 3pi, and the basal energy of a 4-vertex gluon interaction is given by $\dfrac{1}{\sqrt[8]{1 - \dfrac{1}{3\pi}}}$ and is explained below.

Dyson–Schwinger equations, also named Schwinger–Dyson equations, are relations between various Green's functions of a quantum field theory discovered by Freeman Dyson and Julian S. Schwinger. Since they represent the equations of motion for Green's functions, they are also often referred to as the Euler–Lagrange equations of quantum field theory. There are infinitely many functional differential equations, all of which are directly or indirectly coupled to one another. This is why one often speaks of the infinite tower of the Dyson–Schwinger equations.

The Dyson equations were originally derived by Dyson by summing an infinite number of Feynman diagrams and were extended by Schwinger in his quantum action principle to all the Green's functions of any quantum field theory. Graphically, in the Feynman diagrams, swordfish diagrams result if the Dyson–Schwinger equations are applied to the 3 gluon vertices, which are virtual particles. Summing these three half circles, the way length within the 3 gluon vertices is 3 pi in length. If we associate the length of virtual particles with energy, then the energy needed for the 3 gluon vertices is 3 pi as the amount of a normal interaction.

In the following, considerations on partition functions in the context of particles at the elementary level will be presented. These particles consist of either quarks and gluons or only gluons.

Both quarks and leptons are believed to consist of even smaller particles, the rishons. There are two fundamental rishons: the T-rishon and the V-rishon; this is consistent with the H. Harari preon model described, e.g., in "A schematic model of quarks and leptons" Physics Letters 1997 [18].

An important initial hypothesis is that these particles cannot be broken down into even smaller particles.

Another important hypothesis is that fundamental particles (such as V and T) do not possess energy and/or mass. All energy and/or mass is the result of intraparticle interactions.

These basic assumptions indicate that even leptons (e.g., electrons) need to be composed of particles and that an interaction inside leptons is needed.

The total partition function of a particle can be considered an ensemble of subsystems and can be expressed as a product of the individual partition functions for these subsystems:

$$Z = \prod_{j=1}^{N} \zeta_j = \prod_{j=1}^{N} \zeta_{jj}^n = \zeta_2 \cdot \zeta_3 \cdot \zeta_{3-vertex} \cdot \zeta_{4-vertex} = \zeta_{2j}^a \cdot \zeta_{3j}^b \cdot \zeta_{3-vertex,j}^c \cdot \zeta_{4-vertex,j}^d \qquad (10)$$

$$\Omega = \prod_{j=1}^{N} \omega_j = \prod_{j=1}^{N} \omega_{jj}^n = \omega_2 \cdot \omega_3 \cdot \omega_{3-vertex} \cdot \omega_{4-vertex} = \omega_{2j}^a \cdot \omega_{3j}^b \cdot \omega_{3-vertex,j}^c \cdot \omega_{4-vertex,j}^d \qquad (11)$$

Consequently, at the elementary level, the total partition function must result from the product of individual partial partition functions.

### 2.2. Considering Quark- Color Interactions

The task of these individual partial partition functions is to describe the individual interactions in an elementary particle and capture them mathematically. The following question therefore arises: What interactions exist at the elementary level? The answer is the color interactions among three quarks (i.e., rgb and **rgb**). In interactions of this form, three quarks and three standard gluons are always involved simultaneously. The number of standard gluons or interactions is thus three to the power of n. The partial partition function for this type of interaction is consequently three to the power of b, where b is the number of three-gluon interactions that exist per particle.

$$\zeta_3 = \omega_3 = 3^b \qquad \text{for } n=b \text{ three-color interactions per particle} \qquad (12)$$

Another elementary interaction is the color–anticolor interaction between two quarks (e.g., in q$\bar{q}$). Two quarks (a colored quark and an anticolored quark) and two standard gluons (a neutral gluon and a colored gluon) are always involved in this interaction. The number of gluons or interactions is therefore two times the power of a. The corresponding partial partition function is similarly two times the power of a.

Here, a denotes the number of two-gluon interactions per particle. For quantum mechanical reasons, this bond is always mediated by two gluons that are exchanged between the quarks consecutively: a neutral gluon is exchanged, and then, a color gluon is exchanged.

$$\zeta_2 = \omega_2 = 2^a \qquad \text{for } n=a \text{ color–anticolor interaction per particle} \qquad (13)$$

### 2.3 Considering the Gluon- Vertices

According to the Dyson–Schwinger equation (DSE), the **3-vertex** interaction of a gluon can be described as the sum of three semicircles.

The 3-gluon vertex can be converted to a circular structure via DSE on-loop correction. This process creates three circles with an entrance and a 3-vertex, all facing each other at 180° [19]. This corresponds to a path length between the entry and 3 vertices of 3 times 1 pi. When the Dyson–Schwinger equations are used, three such swordfish diagrams arise, each with a weighting of ½, so that the loop correction leads to a total path length of 3 pi for the 3-gluon vertex [19], [20].

$$\zeta_{3-vertex,j} = \omega_{3-vertex,j} = \frac{1}{2}(2\pi) + \frac{1}{2}(2\pi) + \frac{1}{2}(2\pi) = 3\pi \qquad (14)$$

The distance that a gluon travels in this form of interaction is thus 3π times as long as the ordinary gluon distance in one of the two interactions mentioned above. Consequently, the partial partition function for this form of interaction can be described as the term 3 pi to the power of c. Here, c is the number of interactions for this type of particle.

$$\zeta_{3-vertex} = \omega_{3-vertex} = (3\pi)^c \qquad \text{for } n=c \text{ \underline{3-vertex} interactions per particle} \qquad (15)$$

The **4-vertex** interaction between two gluons is considered. The probability of this interaction occurring is proportional to the total probability (100%) minus the probability of the 3-vertex interaction occurring. This probability of the 3-vertex interaction occurring is the reciprocal of the partition function of the 3-vertex interaction. Therefore, the probability of the 4-vertex interaction occurring is proportional to 1-1/(3 pi). Since this interaction corresponds to a collision event involving two of the eight possible gluons, the 8th root of the term 1-1/(3 pi) must be taken. The partial partition function of the 4-vertex interaction is then the reciprocal of this probability and is thus proportional to 1/(8th root(1-1/(3 pi))) to the power of d. Again, d denotes the number of interactions for this type per particle.

$$\zeta_{4-vertex} = \omega_{4-vertex} = \left( \cfrac{1}{\sqrt[8]{1 - \cfrac{1}{3\pi}}} \right)^d \qquad \text{for } n=d \text{ \underline{4-vertex} interactions per particle} \qquad (16)$$

## 2.4 A Closer Look at Quark- Color- Interactions a and b:

### Binding and Nonbinding Events

For the partial partition functions $\zeta_2$ and $\zeta_3$, we need to differentiate between events that have a binding nature, which should be considered negative, and events that have a nonbinding nature, which should be considered positive. This leads us to quotients for $\zeta_2$ and $\zeta_3$, which are defined as follows:

$$\zeta_2 = \omega_2 = \frac{2^{a(nonbinding)}}{2^{a(binding)}} \quad \text{meaning that} \quad a > 0 \Rightarrow nonbinding \quad event \Rightarrow energy \quad needed \qquad (17)$$

$$a < 0 \Rightarrow binding \quad event \Rightarrow energy \quad released$$

$$\zeta_3 = \omega_3 = \frac{3^{b(nonbinding)}}{3^{b(binding)}} \quad \text{meaning that} \quad b > 0 \Rightarrow nonbinding \quad event \Rightarrow energy \quad needed \qquad (18)$$

$$b < 0 \Rightarrow binding \quad event \Rightarrow energy \quad released$$

### 2.5 Calculation of All Possible Combinations of Interaction Possibilities for the Nucleon

### (Proton or Neutron)

To calculate the number of interaction possibilities for the proton, a hypothesis must first be formulated:

The following hypothesis is formulated: in the ground state, only two gluon–gluon interactions can follow one another. More than two interactions can follow one another only in higher or excited energy states of the nucleon. For example, these higher excitation states are realized in the sigma baryon.

On this basis, the number of possible interaction combinations can be counted graphically, as shown in Figure 1. The following number of possible combinations result for a nucleon in the ground state:

The total partition function for the nucleon ((a, b, c, d)=(0, -2, 9, 4)) is as follows:

$$Z_p = \Omega_p = 2^0 \cdot 3^{-2} \cdot (3\pi)^9 \cdot \left( \frac{1}{\sqrt[8]{1-\frac{1}{3\pi}}} \right)^4 = 6.8953127 \cdot 10^7 \qquad (19)$$

or, in logarithmic form:

$$\ln Z_p = \ln \Omega_p = 0 \cdot \ln(2) - 2 \cdot \ln(3) + 9 \cdot \ln(3\pi) - \frac{4}{8}\ln\left(1-\frac{1}{3\pi}\right) \qquad (20)$$

The total partition function for the neutral pion ((a, b, c, d)=(-1, 1, 7, 0)) is as follows:

$$Z_{pion} = \Omega_{pion} = 2^{-1} \cdot 3^1 \cdot (3\pi)^7 \cdot \left( \frac{1}{\sqrt[8]{1-\frac{1}{3\pi}}} \right)^0 = 9.908071 \cdot 10^6 \qquad (21)$$

or, in logarithmic form:

$$\ln Z_{pion} = \ln \Omega_{pion} = -\ln(2) + \ln(3) + 7 \cdot \ln(3\pi) - \frac{0}{8}\ln\left(1-\frac{1}{3\pi}\right) \qquad (22)$$

The total partition function for the neutral sigma baryon ((a, b, c, d)=(-3, 2, 8, 16)) is as follows:

$$Z_{Sigma} = \Omega_{Sigma} = 2^{-3} \cdot 3^2 \cdot (3\pi)^8 \cdot \left( \frac{1}{\sqrt[8]{1-\frac{1}{3\pi}}} \right)^{16} = 8.7648980 \cdot 10^7 \qquad (23)$$

or, in logarithmic form,

$$\ln Z_{Sigma} = \ln \Omega_{Sigma} = -3\ln(2) + 2\ln(3) + 8 \cdot \ln(3\pi) - \frac{16}{8}\ln\left(1 - \frac{1}{3\pi}\right) \qquad (24)$$

For the electron, the following total partition function $Z_e$ is obtained:

$$Z_e = \Omega_e = 2^{-1} \cdot 3^2 \cdot (3\pi)^4 \cdot \left(\frac{1}{\sqrt[8]{1 - \frac{1}{3\pi}}}\right)^4 = 37553 \qquad (25)$$

$$\ln Z_e = \ln \Omega_e = -1 \cdot \ln(2) + 2 \cdot \ln(3) + 4 \cdot \ln(3\pi) - \frac{4}{8}\ln\left(1 - \frac{1}{3\pi}\right) \qquad (26)$$

At first glance, the electron is not associated with gluon interactions. However, according to the Rishons model (or preon model or Harari model), leptons are also composed of electrons from even smaller particles. According to Harari, these are the T- and V-particles. The rishon composition of the electron is anti-T, anti-T, or anti-T. Therefore, the electron is composed of three identical anti-T particles. In a later variant, Harari and Seiberg differentiated the electron to be composed of anti-T-R, anti-T-R, and anti-T-L [21]. They described how the electron is an e$^-$=(T$_R$T$_R$)T$_L$ particle (Table 8, page 156 in [21]).

The meaning of the electron n-tuple (-1, 2, 4, 4) could be interpreted as follows: the electron inside itself has 1 binding 2-particle interaction (the binding hypercolor H according to Harari et al., e.g., anti-T-R and anti-T-R), 2-nonbinding 3-particle interactions (anti-T-R; anti-T-R; anti-T-L and anti-T-L; anti-T-R; anti-T-R), four 3-vertex interactions and four 4-vertex interactions. In the case of the electron interaction, these four 3-vertex interactions and four 4-vertex interactions might be substantially different from the 3-gluon-vertex interactions and 4-gluon-vertex interactions in the case of the more complex particles. Most likely, in the case of the electron, both interactions (3-vertex and 4-vertex) are combined to form four combined interactions (3-vertex/4-vertex). The easiest way to realize this structure is a linear formation. In this linear formation, we start with the first anti-T, followed by the first 3-vertex interaction, the next two 4-vertex interactions, a second 3-vertex interaction, the second anti-T, the third 3-vertex interaction, two more 4-vertex interactions, the fourth 3-vertex interaction and the last anti-T-L. Together, we use the (-1, 2, 4, 4) elements to form the structure.

### 2.6 Elements of the Higher Lepton Generations

Elbaz et al. expanded the Harari preon model toward a higher generation of leptons [22], [23], [24]. They interpreted the higher lepton generations as excited states of the electron.

If we examine the exponent tuples for the muon (-3, 2, 7, 3) and the tauon (1, 2, 7, 7), we observe a further analogy. In the present theory, both are excited states with an increased number of intraparticle interactions. Thus, we obtain the interactions that form the excited states of the leptons and quarks, which Elbaz and others investigated.

A comparison of complex particles and leptons, even if they differ in nature, reveal some type of repetition of the fundamental rules and laws between leptons and more complex particles. Analogously, the interactions and vertices seem to be repeated.

## 2.7 Weak and Strong Interactions

If we associate the n-tuple of compound particles (baryons and mesons) with their strong interaction (color interaction) and the n-tuple of leptons and quarks with their weak interaction, we observe many analogies between strong interactions and weak interactions. Both types of interactions can be further divided into 4 more basal interactions (2-particle, 3-particle, 3-vertex and 4-vertex interactions).

However, we also note some differences between strong interactions and weak interactions. First, we do not see negative a- and/or b-exponents within the weak interaction for quarks, only for leptons.

However, if we transfer this to the beta decay, which is, in most cases, the decay of a neutron, we obtain.

Beta-Decay

$$d^{-0.3} \rightarrow u^{+0.67} + W^- \rightarrow u^{+0.67} + e^- + \underline{v} \quad (27)$$

Rishons:

$$\underline{\mathbf{VVT}}^{-0.3} \rightarrow \mathbf{TTV}^{+0.67} + \underline{\mathbf{(TT)TVV(V)}} \rightarrow \mathbf{TTV}^{+0.67} + \underline{\mathbf{TTT}}^- + \underline{\mathbf{VVV}}$$

Interactions:

$$\underline{(2, 2, 4, 8)} \rightarrow (1, 2, 4, 8) + \underline{(0, 0, 10, 5)} \rightarrow (1, 2, 4, 8) + \underline{(-1, 2, 4, 4)} + \underline{(-1, -2, 0, 0)}$$

Myon-Decay

$$\mu^- \rightarrow W^- + \nu_\mu \rightarrow e^- + \underline{\nu_e} + \nu_\mu \quad (28)$$

Rishon:

$$[\underline{TTT}^- + 3\underline{VV}] \rightarrow \underline{(TT)TVV(V)}^- + VVV \rightarrow \underline{TTT}^- + \underline{VVV} + VVV$$

Interactions:

(-3, 2, 7, 3) → (0, 0, 10, 5) + (-1, -0, 0, 0) → (-1, 2, 4, 4) + (-1, -2, 0, 0) + (-1, -0, 0, 0)

Pion decay

$$\pi^-(\underline{u}, d) \rightarrow W^- \rightarrow \mu^- + \nu_{\underline{\mu}} \quad (29)$$

Rishon:

$$TTV^{-0.67} + VVT^{-0.3} \rightarrow (TT)TVV(V)^- \rightarrow TTT^- + VVV$$

Interactions:

(1, 2, 4, 8) + (2, 2, 4, 8) → (0, 0, 10, 5) → (-3, 2, 7, 3) + (-1;0;0;0)

Strange-Quark-Decay

$$\underline{s}^{+0.33} \rightarrow W^+ + \underline{u}^{-0.67} \rightarrow u^{-0.6} + \pi^+ (u, \underline{d}) \quad (30)$$

from $K^+(\underline{s}, u)$         from $\pi^0 (u, \underline{u})$

Rishon:

$$[2T\underline{T} + V\underline{V} + VVT^{+0.3}] \rightarrow (TT)TVV(V)^+ + \underline{TTV}^{-0.67} \rightarrow TTV^{+0.66} + VVT^{+0.33} + \underline{TTV}^{-0.66}$$

Interactions:

(3, 2, 5, 17) → (0, 0, 10, 5) + (1, 2, 4, 8) → (1, 2, 4, 8) + (2, 2, 4, 8) + (1, 2, 4, 8)

Therefore, the Rishon contours of the strange quark and myon are written according to Elbaz [23,24].

According to the Elbaz notation of the elementary particles [24], which is slightly modified here, the intraparticle interactions are as follows:

Electron: a=-1[**TTT**(V**V**)]    Myon: a=-3 [**TTT**(V**V**)$_3$]    strange-Quark: a=3 [**VVT**(T**T**)$_2$(**VV**)]

$a=4-1=3$

[**V** V ….**T**    (-1, 2, 4, 4)    [V V…**T**   (-3, 2, 7, 3)    [**V** V …**V**    (3, 2, 5, 17)

    **T**                                  **V** V… **T**                              T T… ..**T**

    **T** ]                                **V** V…**T**]                             **V** ….T **T**  ]

1 binding                       3 binding                          4 nonbinding, 1 binding

It can be recognized that a binding 2-particle interaction (a<0) occurs between an antimatter T and a matter V and/or a matter T and an anti-matter V. It seems that **both need to change** T to V and matter to antimatter to form a binding interaction.

On the other hand, a matter V-to-antimatter V interaction and/or a matter T-to-antimatter T interaction, as in the strange quark, is nonbinding (a>0). Additionally, T-to-V binding and/or anti-T-to-anti-V binding, such as in the u-quark, d-quark, and s-quark, is nonbinding. Homomorphic binding, such as TT, **TT** and/or VV and/or **VV** binding, in the particles is not considered.

u-quark a=1 [TTV]                u-quark  a=1 [**TTV**]               d-quark   a=2 [**VTV**]

[       T        (1,2,4,8)        [**T**    (1,2,4,8)               [ **V**          (2,2,4,8)

    T                                  **T**                                  T

    V ]                                **V** ]                                V ]

1 nonbinding                    1 nonbinding                      2 nonbinding

Tauon: a= 1[**TTT**(V**V**) $_6$]     W$^+$-particle a=0 [TTT(VV**V**)]      Z-particle: a=0 [**VVV**(VVV)]

[ ------**V** V ….**T**   (1,2,7,7)     [ V…T   (0,0,10,5)               [ V … **V**(0,0,10,14)

    **V** V… **T**                         V… T                                V…. **V**

    **V** V….. **T**                       V…T]                               V…. **V**  ]

    (V **V**)$_3$

1 nonbinding                    zero binding                      zero binding

ν_τ=[VVV(VV)₃]                     W⁻-particle a=0 [**<u>TTT(VVV)</u>**]

ν_e=[VVV]

When the W particle is formed, TTV is added to d, and the a and b interactions are somewhat lost. Only the c and d interactions are possible in the W particle. Next, the W particle forms two particles with a and b interactions again.

The Elbaz notation was modified to meet the particle decay requirements.

<u>For the proportionality factor between the value of the partition function and the rest energy (equivalent to the rest mass) of the particle, the following values are obtained for the nucleon, electron and sigma baryon:</u>

<u>for the proton:</u>

$$f_p = E_{0,p} / Z_p = \frac{E_{0,p}}{\Omega_p} = 938.272 \cdot 10^6 eV / 6.8953127 \cdot 10^7 = 13.6074137 eV \quad (31)$$

<u>for the electron:</u>

$$f_e = E_{0,e} / Z_e = \frac{E_{0,e}}{\Omega_e} = 0.51099895 MeV / 37553 = 13.6074068 eV \quad (32)$$

<u>for the sigma baryon:</u>

$$f_{Sigma} = E_{0,\Sigma} / Z_\Sigma = \frac{E_{0,\Sigma}}{\Omega_\Sigma} = 1192.642 \cdot 10^6 eV / 8.7648980 \cdot 10^7 = 13.60702657 eV \quad (33)$$

## 2.8 Proportionality between Partition Function and Potential Energy in Closed Loop Diagrams:

### The Rydberg Proportionality Factor

This factor is almost identical to the Rydberg energy ($E_{Ryd}$=13.605693122 eV). At first glance, this result seems astonishing. Since the 1s electron in the hydrogen atom is considered almost electrostatically bound and/or QED--bound to the proton, we might expect that the Rydberg energy would be related to electrostatic and/or QED binding. If we assume, however, that the Rydberg energy does not merely represent the binding energy of the electron to the proton but rather should be considered the smallest energy quantum for all binding, then the problem is solved. This smallest amount of binding represents binding in general and the binding energy in every particle. Additionally, the greater the number of interactions presents in a particle and the greater the number of binding-quants that are therefore needed and present in the particle, the greater the binding energy that must consequently result and be present in the corresponding particle. This binding energy now seems to exist in a quantized form only, and the smallest quantum of this binding energy seems to be identical to the Rydberg energy. Hence, the Rydberg energy is not only the ionization energy of the hydrogen atom but also has a much more general meaning and importance.

If we consider the zero-tuple (0, 0, 0, 0), then this zero-tuple has a value of 1 Rydberg. Therefore, for binding mediated through zero gluons and with zero vertices involved, a binding energy of 1 Rydberg remains.

It could also be that, in a yet-to-be developed theory that is not within the well-established theories of QCD and/or QED, the binding of the 1s electron to the proton in hydrogen and helium is mediated through a new quantum represented by the zero-tuple (0, 0, 0, 0) (referred to as (0, 0, 0, 0)- binding herein); this (0, 0, 0, 0) binding and the other particular binding energies seem to have a common origin, as they are similar in nature and energy level. They represent an intraparticle quantum–mechanical binding state. In this newly developed theory, the binding of the 1s electron would be not only electrostatic and/or QED in nature but also novel in terms of its general quantum-mechanical nature and would probably be mediated by a new quantum particle. This could be the case because of the close distance between the 1s electron and the proton and should then be related to the 1s electron only.

For quantum mechanical reasons, the density distribution of the 1s electron can approach that of the proton extremely closely.

The Rydberg energy seems to be the amount of energy of one microstate of the microcanonical ensemble that is formed by the binding states of an elementary particle. The partition function Z given here yields the number of microstates realized in one elementary particle, which can be visualized as an ensemble of Feynman-like structures that describe the elementary particle in question.

However, neutrinos should have negative exponents, which would allow for smaller binding energies than Rydberg's energy. The experimental results suggest a neutrino mass smaller than <1 eV/c2.

**2.9 Short Summary of the Relations Novel Found**

We draw the following conclusions, which are valid for particles only:

$$m \sim E \sim Z \sim \Omega \sim N_{microstates} \quad \text{or} \quad m \sim E \sim \Omega \tag{34a+34b}$$

$$\frac{E}{Z} = E_{Rydberg} \quad \text{or} \quad \frac{E}{\Omega} = E_{Rydberg} \tag{35a+35b}$$

$$E = E_{Rydberg} \cdot Z \quad \text{or} \quad E = E_{Rydberg} \cdot \Omega \tag{36a+36b}$$

$$E = E_{Rydberg} \cdot e^{\ln Z} \quad \text{or} \quad E = E_{Rydberg} \cdot e^{\ln \Omega} \tag{37a+37b}$$

**3.0 Discussion**

**3.1 Generalization of the partition function and generalization of the energy (mass) relation**

A generalization of the partition function yields the following general formulation of the partition function of a particle.

The mass of an elementary particle can also be expressed in general via the following relationship.

$$E_{particle} = m_{particle} c^2 = 2^a \cdot 3^b \cdot (3\pi)^c \cdot \left( \frac{1}{\sqrt[8]{1 - \frac{1}{3\pi}}} \right)^d \cdot E_{Ryd} \tag{38 similar to (6)}$$

Here, the exponents a, b, c, and d are related to the number of interactions present in the particle.

This formulation yields the following hypotheses:

**1.** The constant of proportionality between the partition function of a particle and the energy (mass) of that particle is always the Rydberg energy (energy per microstate).

$$Z = \Omega = 2^a \cdot 3^b \cdot (3\pi)^c \cdot \left(\frac{1}{\sqrt[8]{1-\frac{1}{3\pi}}}\right)^d \text{ and } \ln Z = a \cdot \ln(2) + b \cdot \ln(3) + c \cdot \ln(3\pi) - \frac{d}{8}\ln\left(1 - \frac{1}{3\pi}\right)$$

(39)

$$E = 2^a \cdot 3^b \cdot (3\pi)^c \cdot \left(\frac{1}{\sqrt[8]{1-\frac{1}{3\pi}}}\right)^d \cdot E_{Ryd}$$

(40)

**2.** The individual exponents a, b, c, and d are always small whole numbers that represent the number of interactions. Z describes the number of possible combinations of individual interactions. For all known particles, these exponents (a, b, c, d) can be determined through trial and error, which leads us to the exponential series given in Table 1. These values are obtained by comparing the masses for well-known particles resulting from the formula given above with the experimentally known values from the Particle Data Group. Another approach for determining the number of possible combinations consists of the graphic representation of all conceivable combination possibilities, analogous to what has been demonstrated here for the proton or nucleon and the pion.

**3.)** A closer examination of the values in Table 1 reveals that some particles have no quark–quark interactions. These particles seem to exist only because of the presence of gluons; examples include W, Z and H particles. Because these particles are bosons and are known to transmit forces, this finding is consistent with existing knowledge about particles.

**4.)** Experimentally determined energy states of excited sigma baryons can be described with a q**q** model, which is consistent with the description used here (quark–diquark model). The quark–diquark model was developed by Lichtenberg et al. (1968) [25] and has been successfully used to describe hadron spectroscopy (Nagata et al. [26]). The model for the sigma baryon described herein is very similar to the quark–diquark model and can potentially explain why this model works very well. The exponential series for the delta-resonances (uud and ddu) are (-3, 2, 8, 18). Like the sigma baryon, the delta resonances also have an a of -3. Therefore, this could explain why the quark–diquark model of Lichtenberg also fits very well for the delta and nucleon resonances.

### 3.2 Comparison with textbook knowledge

In textbooks [17], a proportional relation between energies and/or between the potency of energy and the partition function is described for a microcanonical ensemble. This relation is different than that for a canonical or macrocanonical ensemble. For those two types of ensembles, a logarithmic relationship between the energy and the partition function has been described. Because we are instead analyzing a microcanonical ensemble and the microcanonical partition function, it could be expected that a linear relationship exists between the appropriate partition function and the binding energy.

### 3.3 Considerations of the accuracy of the calculations presented here

To compare the accuracy of the calculations presented here with that of the previously available formulas, we determine the electron mass or rest energy in two ways.

On the one hand, we calculate the electron rest energy via the new method presented here, and on the other hand, we use the well-known (and similar) relationship between the Rydberg constant and the electron mass (Ry=13.605693122 eV).

$$E_{e,0} = \frac{1}{2} \cdot 3^2 \cdot (3\pi)^4 \cdot \left( \frac{1}{\sqrt[8]{1 - \frac{1}{3\pi}}} \right)^4 \cdot E_{Ryd} = 37554 \quad Ry \tag{41}$$

We compare the results of both methods for determining the electron mass. With the new method, the rest energy of the electron, measured in Rydbergs as the energy unit, is 37554 Ry (nominal experimental value: 37557 Ry); this value is accurate to four decimal places (1 $E_{Ryd}$=13.605693122 eV and $m_{Ryd}$=13.605693122eV/$c^2$).

$$m_e = E_{Ryd} \frac{\varepsilon^2 h^2 8}{e^4} = 9.11282233567 \cdot 10^{-31} \quad kg \tag{42}$$

According to the analogous previously well-known common relationship, the electron mass is 9.112822335674E-31 kg (nominal experimental value of 9.109 E-31 kg), which corresponds to an accuracy of only two decimal places. The relationship between the Rydberg constant and the electron mass presented here is therefore far more precise than the previously known relationship, with an accuracy that is higher by two decimal places (a factor of approximately 100).

**4.4 Competing Interests Statement**

The author declares that he has no competing interests. The author declares that he has no known competing financial interests or personal relationships that could have appeared to influence the work reported.

## 6.0 Figure Legends

### 6.1 Figure 1

Different types of interactions occur within a nucleon: color–anticolor interactions, which are 2-particle interactions; normal color interactions, which are three-particle interactions; 3-gluon vertices; and 4-gluon vertices. These interactions can also be graphically displayed analogously to the chemical bonding in molecules, as shown here for the nucleon.

a.) Graphical representation of the number of color–anticolor interactions (a) for the nucleon b.) Graphical representation of the number of three-color interactions (b) for the nucleon c.) Graphical representation of the number of 3-gluon vertex interactions (c) for the nucleon d.) Graphical representation of the number of 4-gluon vertex interactions (d) for the nucleon. Therefore, 3-gluon vertex interactions and 4-gluon vertex interactions are combined.

### 6.2 Figure 2

Different types of interactions occur within a neutral pion: color–anticolor interactions, which are 2-particle interactions; normal color interactions, which are normal three-particle interactions; 3-gluon vertices; and 4-gluon interactions. These interactions can be displayed graphically analogously to the chemical bonding in molecules, as shown here for the neutral pion, which was actually a pion–pion dimer.

a.) Graphical representation of the number of color–anticolor interactions (a) for the neutral pion. The neutral pion is considered a quantum-mechanical state of superposition. This superposition is considered to form a pion dimer, suggesting that both pions involved enter into interactions.

b.) Graphical representation of the number of three-color interactions (b) for the pion dimer interaction c.) Graphical representation of the number of 3-vertex gluon interactions (c) for the pi dimer d.) Graphical representation of the number of 4-vertex gluon interactions (d) for the pion dimer. Again, 3-gluon vertex interactions and 4-gluon vertex interactions are combined.

Since the pion is considered a superposition and/or dimer, the numbers a, b, c, and d need to be divided by two.

## 6.3 Figure 3

Different types of interactions occur within a sigma baryon: color–anticolor interactions, which are 2-particle interactions; normal color interactions, which are normally three-particle interactions; 3-gluon vertices; and 4-gluon interactions. These interactions can be displayed graphically analogously to the chemical bonding in molecules, as shown here for the sigma baryon. The sigma baryon carries a strange quark, which has the ability to participate in two particle interactions (=color–anticolor interactions).

a.) Graphical representation of the number of color–anticolor interactions (a) for the sigma baryon

The strange quark seems to have the ability to interact in anticolor interactions; therefore, it seems to have the ability to carry anticolor.

b.) Graphical representation of the number of three-color interactions (b) for the sigma baryon c.) Graphical representation of the number of 3-vertex gluon interactions (c) for the sigma baryon d.) Graphical representation of the number of 4-vertex gluon interactions (d) for the sigma baryon. Again, 3-gluon vertex interactions and 4-gluon vertex interactions are combined. Two 4-gluon interactions seem to occur subsequently in the sigma baryon.

# 7.0 Tables

## 7.1 Table 1

This table lists the exponent series (a, b, c, and d) for the most important particles.

**Figure 1** Protons duu and Neutrons udd (a=0; b=-2; c=9; d=4)
Quantum State Sum Ensembles for the Parameters a and b
in the protons and/or neutrons we have the following interactions:

*a.)*

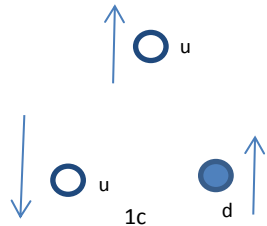

1c

no color– anticolor interaction
a=0

*b.)*

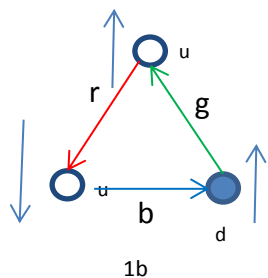

1b

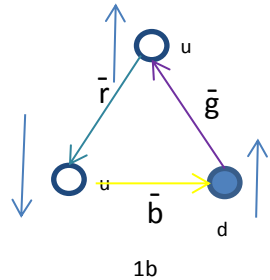

1b

one three color interaction rgb
one three anticolor interaction
b=-2
both are binding

**Figure 1** Protons and Neutrons (a=0; b=-2; c=9; d=4)
Quantum State Sum Ensembles for the Parameters c and d
3-gluon vertices (c) and 4-gluon vertices (d)

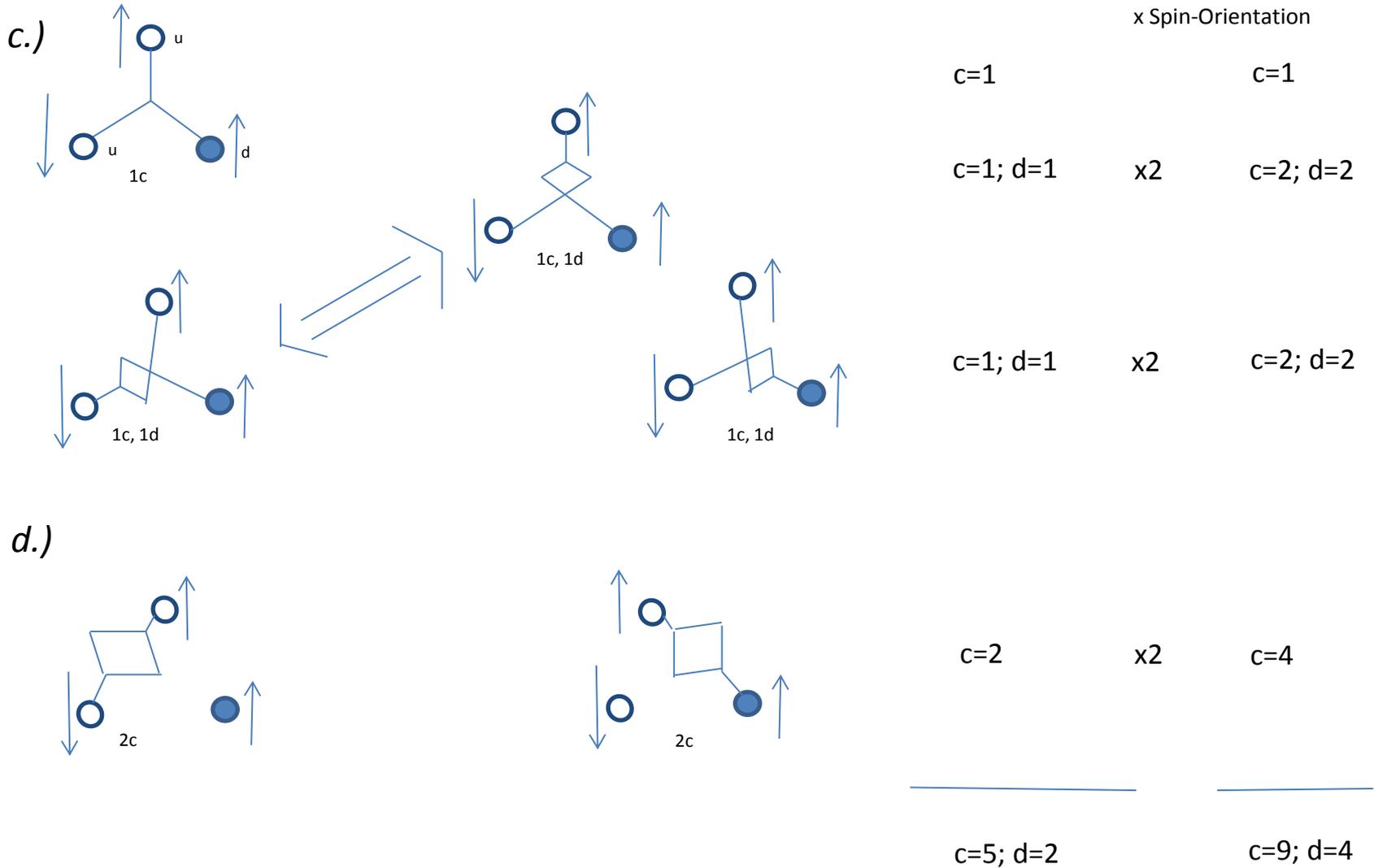

**Figure 2** neutral Pions  uū+dd̄ (a=-1; b=1; c=7; d=0)
Quantum State Sum Ensembles for the Parameters a and b
formulated as pion-dimers analog to the wave function

*a.)*

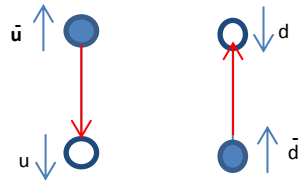

two color– anticolor interactions

a=2 for dimer
a=-1 per monomer - binding

*b.)*

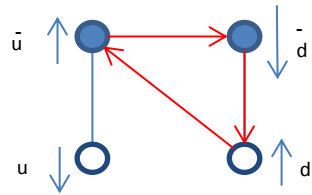

two three color interactions rgb

b=2 for dimer
b=1 per monomer - nonbinding

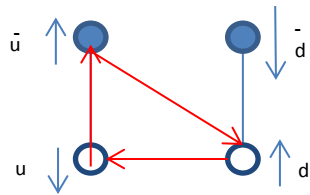

**Figure 2** neutral Pions (a=-1; b=1; c=7; d=0) State Sum, Parameters c and d

*c.)*

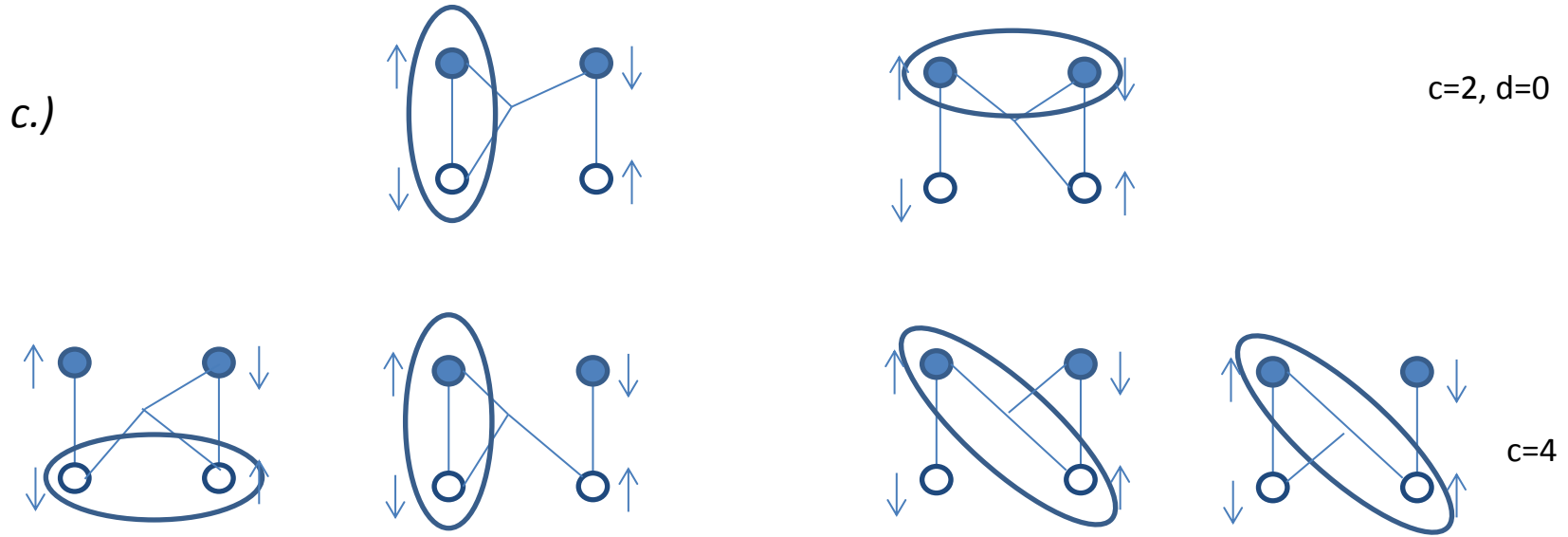

c=2, d=0

c=4

*d.)*

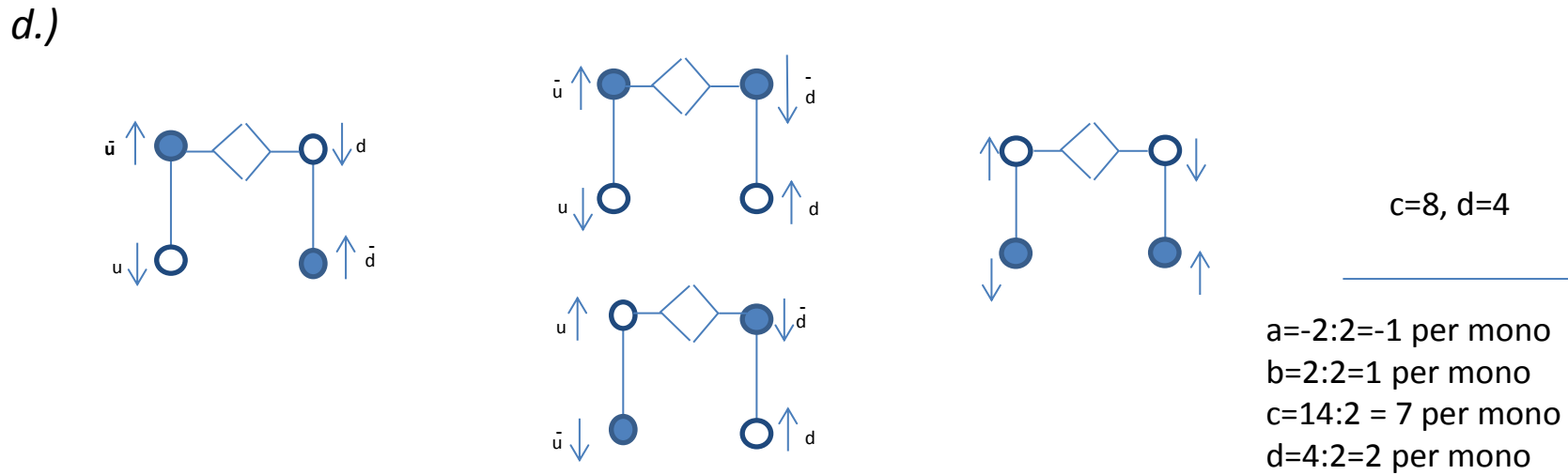

c=8, d=4

---

a=-2:2=-1 per mono
b=2:2=1 per mono
c=14:2 = 7 per mono
d=4:2=2 per mono

**Figure 3** neutral Sigma-baryon uds (a=-3; b=2; c=8; d=16) Parameters a and b

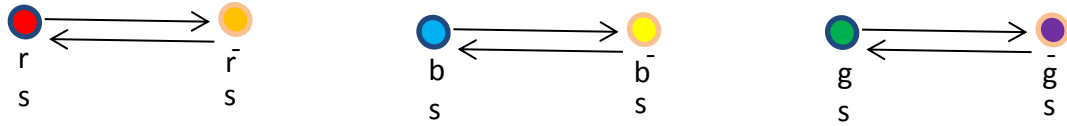

*a.)*

three color– anticolor interactions

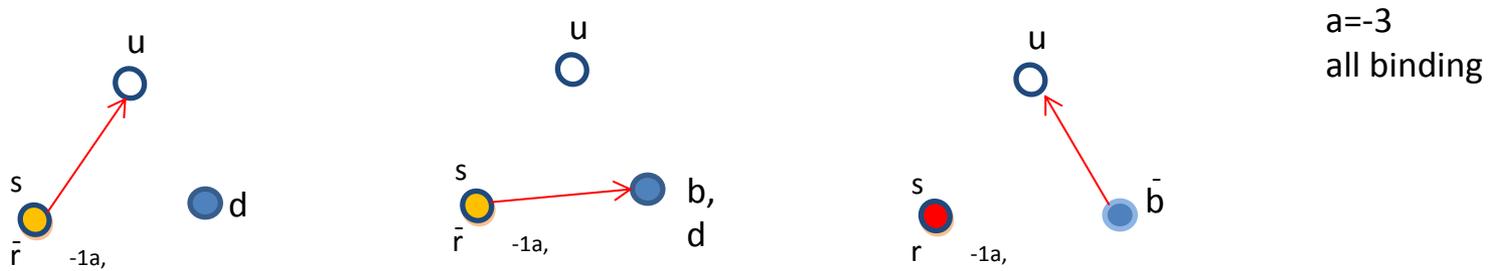

a=-3
all binding

s-quark participates in anticolor interaction

*b.)*

two three color interactions rgb

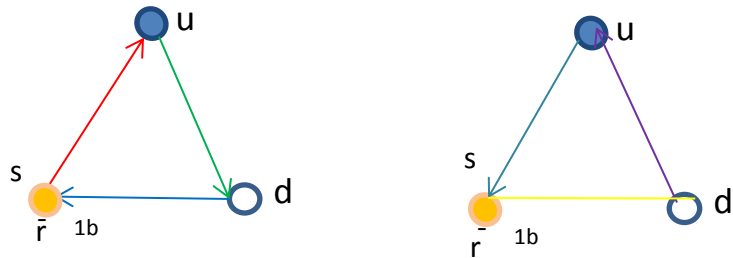

b=2
nonbinding

**Figure 3** neutral Sigma-baryon (a=-3; b=2; c=8; d=16) Parameters c and d

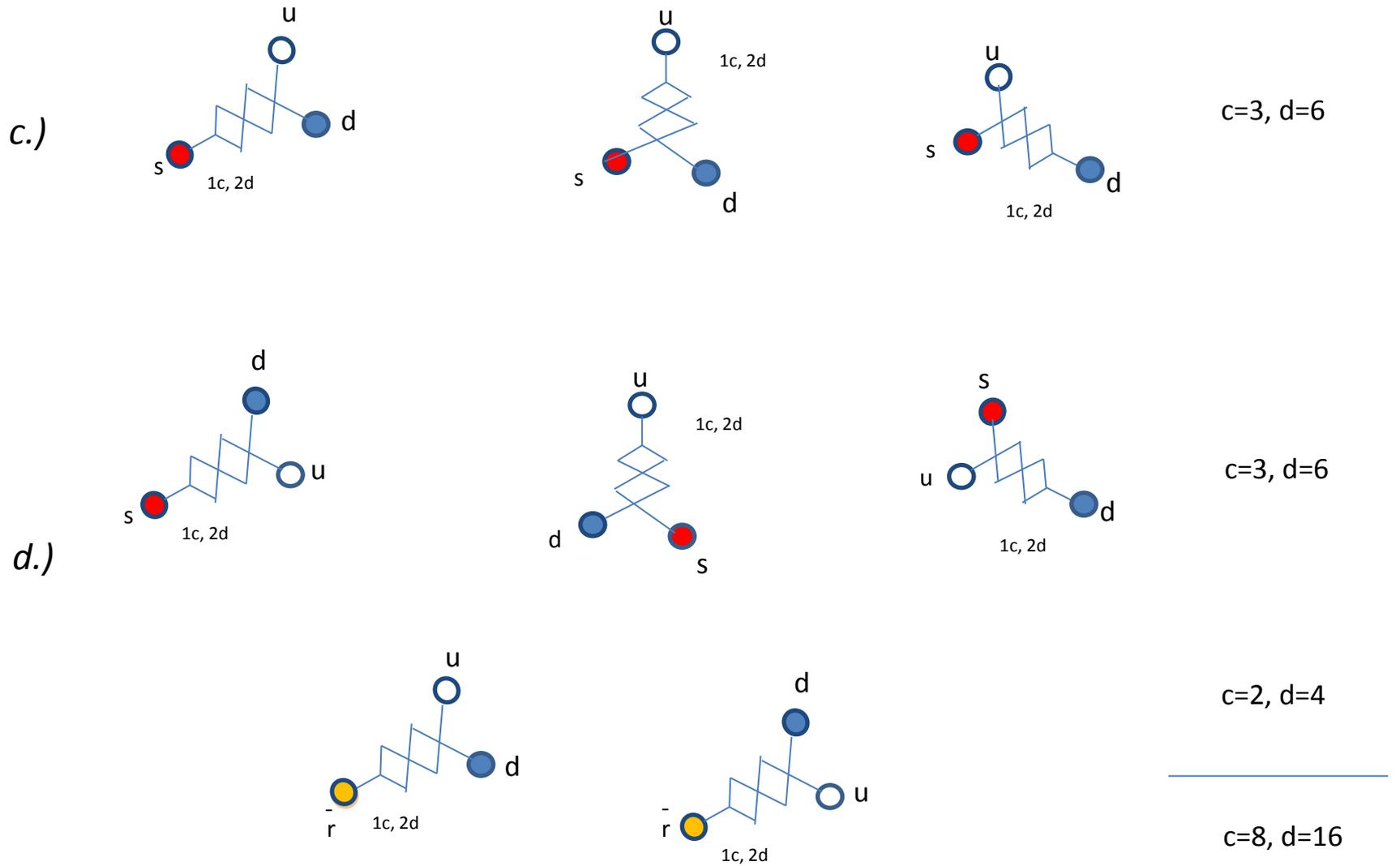

**Table 1   Exponent- series of the mass- formula calculated for the different particles**

| a | b | c | d | mass/[MeV] | particle | composition |
|---|---|---|---|---|---|---|
| **leptons** | | | | | | |
| -1 | -2 | 0 | 0 | 7E-7 | neutrino | VVV |
| -1 | 2 | 4 | 4 | 0.5109 | electron | (TT)T |
| -3 | 2 | 7 | 3 | 105.44 | muon | TTT* |
| 1 | 2 | 7 | 7 | 1784.49 | tauon | TTT** |
| **quarks** | | | | | | |
| 1 | 2 | 4 | 8 | 2.16 | up | u,(TT)V, TVT, VTT |
| 2 | 2 | 4 | 8 | 4.67 | down | d, (vv)t, tvt, tvv |
| 2 | 3 | 6 | 15 | 1270 | charm | c TVT* |
| 3 | 2 | 5 | 17 | 92 | strange | s tvv* |
| 1 | 2 | 9 | 13 | 172421 | top | t TVT** |
| 2 | 3 | 6 | 100 | 4184 | bottom | b tvv** |
| **vector bosons** | | | | | | |
| 0 | 0 | 10 | 5 | 80700.0 | W | |
| 0 | 0 | 10 | 14 | 91554.0 | Z | |
| 0 | 0 | 10 | 36 | 124634.0 | H | |
| **mesons** | | | | | | |
| -1 | 1 | 7 | 0 | 134.8 | pi 0 | (uu-dd)/2 |
| -1 | 0 | 8 | 11 | 494128.0 | K+ | us |
| -1 | -2 | 9 | 15 | 547299.0 | eta 0 | uu+dd-2ss |
| 0 | 0 | 8 | 9 | 960.9 | eta dash | uu+dd+ss |
| 1 | 0 | 8 | 7 | 1868.71 | D+ | cd |
| 1 | 0 | 8 | 11 | 1976.5 | Ds+ | cs |
| -1 | 0 | 9 | 20 | 5283.0 | B meson | ub |
| -1 | 0 | 9 | 22 | 5433.0 | strange meson | sb |
| -1 | 0 | 9 | 33 | 6339.0 | charm meson | cb |
| **vector mesons** | | | | | | |
| 3 | 0 | 7 | 5 | 771.0 | rho 0 | (uu-dd)/2 |
| 0 | 0 | 8 | 4 | 895872.0 | K*0 | ds |
| 3 | 0 | 7 | 6 | 782.0 | omega | (uu+dd)/2 |
| 0 | 0 | 8 | 13 | 1016 | phi | ss |
| 1 | 0 | 8 | 12 | 2004.0 | D*0 | cu |
| 1 | 0 | 8 | 43 | 3095.6 | J/psi | cc |
| **baryons** | | | | | | |
| 0 | -2 | 9 | 4 | 938.15 | proton | uud |
| -3 | 2 | 8 | 11 | 1111.0 | lambda | uds |
| -3 | 2 | 8 | 16 | 1192.5 | sigma 0 | uds |
| -3 | 2 | 8 | 18 | 1226.4 | delta | ddd |
| -3 | 2 | 8 | 23 | 1315.5 | xi 0 | uss |
| -1 | -1 | 9 | 3 | 1387.6 | sigma-* | dds |
| -3 | 2 | 8 | 29 | 1430.95 | N(1440) | udd |
| -1 | -1 | 9 | 10 | 1530.7 | xi 0 reson | uss |
| 1 | 0 | 8 | 1 | 1670.0 | Omega - | sss |

| -3 | 2 | 8 | 63 | 2304.0 | lambda-c | udc |
| --- | --- | --- | --- | --- | --- | --- |
| -3 | 2 | 8 | 68 | 2472.0 | c-sigma | ddc |
| | | | 129 | 5815.0 | cascade B | usb |
| -3 | 2 | 8 | 130 | 5897.0 | bottom sigma | ddb |
| -3 | 2 | 8 | 72 | 2614.9 | charmed xi prime | usc |
| -3 | 2 | 8 | 96 | 3661.0 | double charmed xi | ucc |